\documentclass[twocolumn]{aastex63}
\usepackage{color}
\usepackage[titletoc]{appendix}
\usepackage{amsmath}
\usepackage{amssymb}
\usepackage{mathtools}
\usepackage{upgreek}
\usepackage{float}
\usepackage{comment}
\usepackage{enumitem}
\usepackage{natbib}
\usepackage{graphicx}
\usepackage{bm}
\usepackage{totcount}
\usepackage{multirow}

\newtotcounter{citnum} %From the package documentation
\def\oldbibitem{} \let\oldbibitem=\bibitem
\def\bibitem{\stepcounter{citnum}\oldbibitem}

\shortauthors{Millholland \& Winn}
\shorttitle{Split Peas in a Pod}

\begin{document} 

\title{Split Peas in a Pod: \\ Intra-System Uniformity of Super-Earths and Sub-Neptunes}

\author[0000-0003-3130-2282]{Sarah C. Millholland}
\altaffiliation{NASA Sagan Fellow}
\affiliation{Department of Astrophysical Sciences, Princeton University, 4 Ivy Lane, Princeton, NJ 08544, USA}
\email{sarah.millholland@princeton.edu}

\author[0000-0002-4265-047X]{Joshua N.\ Winn}
\affiliation{Department of Astrophysical Sciences, Princeton University, 4 Ivy Lane, Princeton, NJ 08544, USA}

\begin{abstract}
The planets within compact multi-planet systems tend to have similar sizes, masses, and orbital period ratios, like ``peas in a pod''. This pattern was detected when considering planets with radii between 1 and 4\,$R_\oplus$. However, these same planets show a bimodal radius distribution, with few planets between 1.5 and 2\,$R_{\oplus}$. The smaller ``super-Earths'' are consistent with being stripped rocky cores, while the larger ``sub-Neptunes'' likely have gaseous H/He envelopes. Given these distinct structures, it is worthwhile to test for intra-system uniformity separately within each category of planets. Here, we find that the tendency for intra-system uniformity is twice as strong when considering planets within the same size category than it is when combining all planets together. The sub-Neptunes tend to be $1.7^{+0.6}_{-0.3}$ times larger than the super-Earths in the same system, corresponding to an envelope mass fraction of about 2.6\% for a $5\,M_{\oplus}$ planet. For the sub-Neptunes, the low-metallicity stars are found to have planets with more equal sizes, with modest statistical significance ($p\sim 0.005$). There is also a modest ($\sim$2-$\sigma$) tendency for wider-orbiting planets to be larger, even within the same size category.
\\
\end{abstract}

\section{Introduction}
\label{sec: Introduction}

The NASA Kepler mission found hundreds of multiple-planet systems in which the orbital periods are all shorter than 100 days and the planets' sizes are in between those of Earth and Neptune \citep{2011ApJS..197....8L, 2014ApJ...784...44L, 2014ApJ...784...45R, 2014ApJ...790..146F}. Within a given system, the sizes, masses, and orbital period ratios of the planets tend to be more alike than the properties of planets drawn from different systems \citep{2018AJ....155...48W, 2017ApJ...849L..33M}. These intra-system regularities have been nicknamed the ``peas-in-a-pod pattern'' (see the forthcoming review by \citealt{Weiss2021}). At nearly the same time, it was recognized that the radius distribution of these planets is bimodal, with few planets having sizes between 1.5 and 2\,$R_{\oplus}$ \citep{2017AJ....154..109F, 2018MNRAS.479.4786V, 2018AJ....156..264F}. This so-called ``radius valley'' had been predicted in planetary atmosphere models \citep{2013ApJ...775..105O, 2013ApJ...776....2L, 2014ApJ...795...65J} as an effect of photoevaporation, which turns out to be a threshold process leading either to a ``super-Earth'' completely stripped of its gaseous envelope or a ``sub-Neptune'' with enough gas to inflate its transit radius by about a factor of two. Another possibility for these two distinct planet categories is atmospheric loss due to the heat of formation of the solid core \citep{2018MNRAS.476..759G, 2019MNRAS.487...24G, 2020MNRAS.493..792G}.

At first glance, these observations appear to be contradictory. Can these systems really be considered peas in a pod when their constituent planets fall into two disparate categories? One possible resolution is that a given star tends to host either mostly super-Earths or mostly sub-Neptunes; however, this is only partially the case, as demonstrated below. We have found that another important part of the resolution is that the intra-system size uniformity within each planet category is so strong that it can be detected even when the two categories are combined together before performing statistical tests. In this sense, intra-system uniformity has been underestimated in previous studies. Although super-Earths and sub-Neptunes might originate from the same initial population of gas-enshrouded rocky planets \citep{2017ApJ...847...29O, 2020AJ....160..108B, 2021MNRAS.503.1526R}, today they have distinct physical structures. In this Letter, we take advantage of our knowledge of these differing structures to revisit the issue of intra-system uniformity within individual planetary categories.

\section{Planet Sample}
\label{sec: planet sample}

We begin with the Kepler planet sample constructed by \cite{2020AJ....160..108B}, which was based on cross-matching the Kepler Object of Interest (KOI) catalog from the NASA Exoplanet Archive with the Gaia-Kepler Stellar Properties Catalog of \cite{2020AJ....159..280B}. The stellar parameters were homogeneously derived using isochrones and broadband photometry, Gaia Data Release 2 parallaxes \citep{2018A&A...616A...1G}, and spectroscopic metallicities whenever they were available. After a series of quality cuts, the sample consisted of 2956 stars hosting 3898 planets. The median fractional uncertainty in the planet radius is 7\%.

\begin{figure}
\centering
\epsscale{1.1}
\plotone{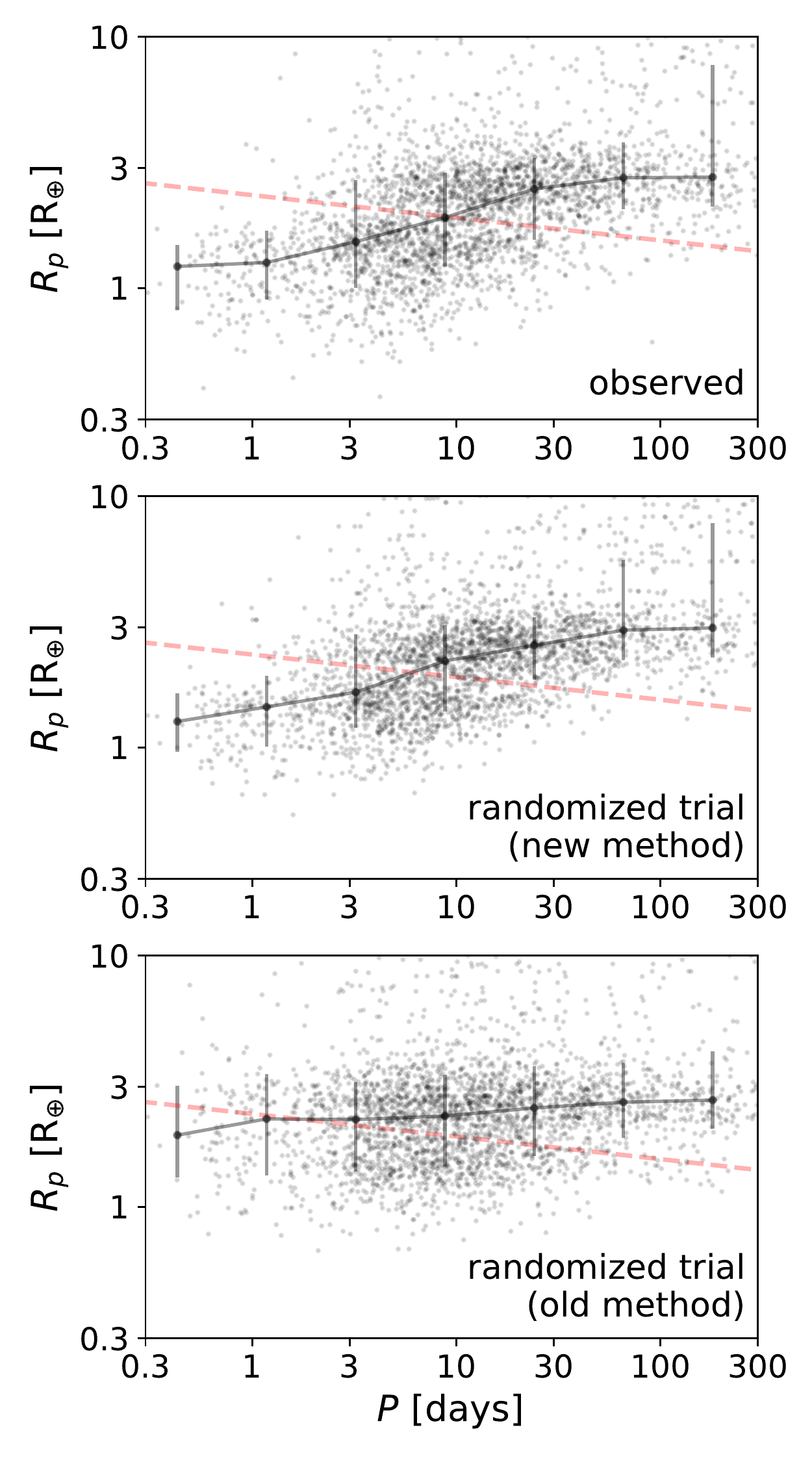}
\caption{Planet sample in period-radius ($P$--$R_p$) space. Top panel: The observed planet sample described in Section~\ref{sec: planet sample}. Middle panel: One randomized trial of the population using our randomization procedure intended to preserve structure in $P$--$R_p$ space (the radius valley and the sub-Neptune desert). Bottom panel: One randomized trial of the population using a procedure that assigns radii independently from periods. In all panels, the red line is the radius valley boundary derived by \cite{2018MNRAS.479.4786V}, and the black line represents the median and 16th and 84th percentiles of the radius within logarithmic bins of period.} 
\label{fig: P-Rp space}
\end{figure}

We consider only planets smaller than $16\,R_{\oplus}$ with orbital periods shorter than 300\,days. We also apply some additional quality cuts. To avoid large systematic errors in planet radii, we discard stars for which \cite{2017AJ....153...71F} found a companion star that contributed more than 5\% of the light in the Kepler photometric aperture. We remove planets with fractional radius uncertainties greater than 50\%. Following the advice of \citet{2020AJ....160...89P}, we omit planets for which the transit duration $T$ was shorter than half of the maximum possible transit duration for a circular orbit,
\begin{equation}
\begin{split}
T_0\equiv \left[\frac{3}{\pi^2G}\frac{P}{\rho_{\star}}\right]^{1/3}=13\,\mathrm{hr}\left(\frac{P}{\mathrm{yr}}\right)^{1/3}\left(\frac{\rho_{\star}}{\rho_{\odot}}\right)^{-1/3}.
\end{split}
\label{eq: T0}
\end{equation}
This is because low values of $T/T_0$ usually imply nearly-grazing transits for which the planet radius is especially uncertain. Finally, we remove planets for which the calculated transit signal-to-noise ratio (SNR) was less than 7.\footnote{To verify that these cuts do not introduce biases, we repeated all calculations in the paper using weaker thresholds, $T/T_0>0.2$ (rather than $T/T_0>0.5$) and SNR $>5$ (rather than SNR $>7$). We find very similar results for all calculations, and the major findings are unchanged.} The SNR is calculated as \citep{2012PASP..124.1279C}
\begin{equation}
\mathrm{SNR}=\frac{(R_p/R_{\star})^2}{\mathrm{CDPP}_{\mathrm{eff}}}\sqrt{\frac{t_{\mathrm{obs}}f_0}{P}},
\end{equation}
where $t_{\mathrm{obs}}$ is the time interval over which data were collected ($\sim$4 years in most cases), $f_0$ is the duty cycle, and $\mathrm{CDPP}_{\mathrm{eff}}$ is the effective combined differential photometric precision,
\begin{equation}
\mathrm{CDPP}_{\mathrm{eff}}=\mathrm{CDPP}_{6\,\mathrm{hr}}\sqrt{\frac{6\,\mathrm{hr}}{T}}.
\end{equation}
For each target star, we obtained $t_{\mathrm{obs}}$, $f_0$, and $\mathrm{CDPP}_{6\,\mathrm{hr}}$ from the Kepler Q1-Q17 stellar properties catalog \citep{2017ApJS..229...30M}. Our final sample contains 2111 stars hosting 2883 planets, with 520 of the systems hosting two or more planets. Figure \ref{fig: P-Rp space} (top panel)
shows the orbital periods and planet radii.

\begin{figure}[t!]
\epsscale{1.1}
\plotone{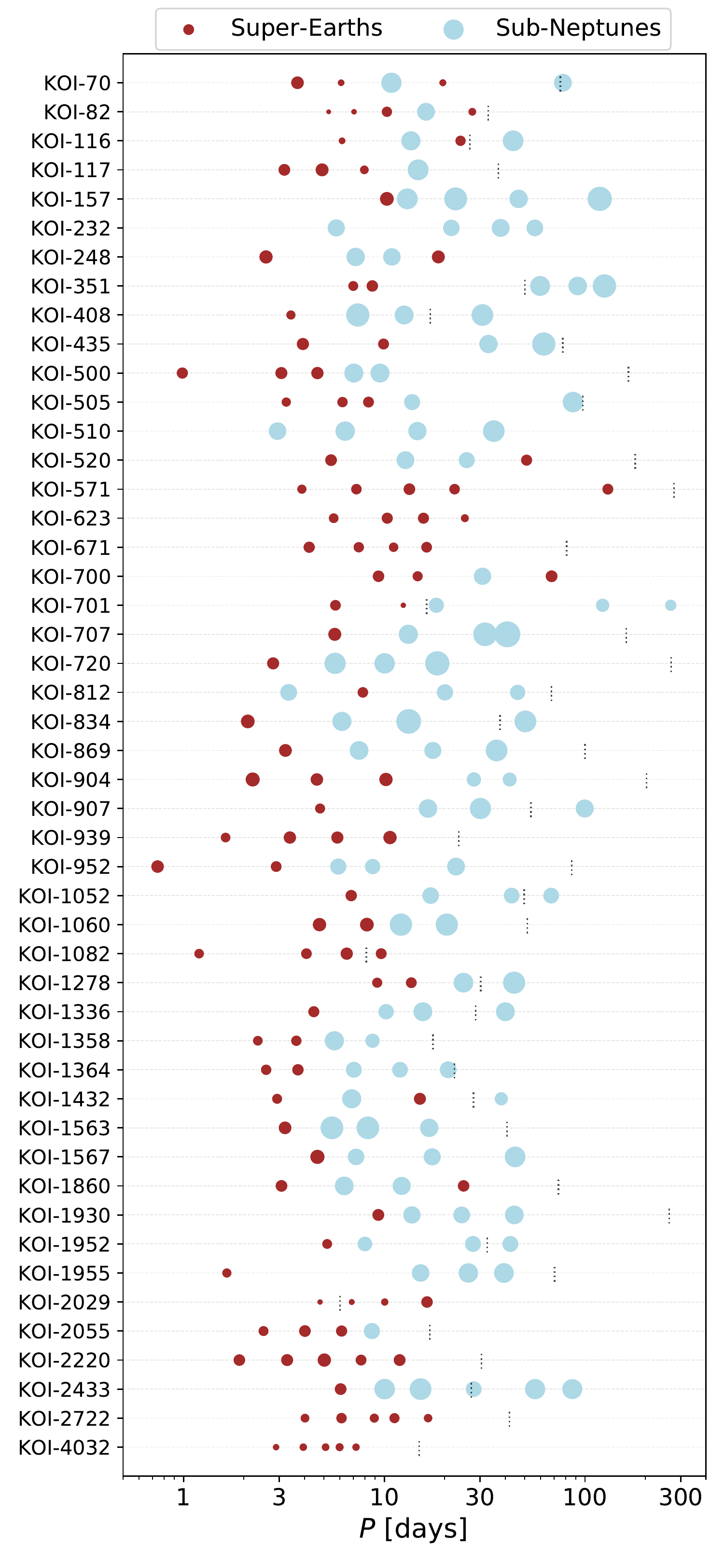}
\caption{Architectures of systems with four or more transiting planets smaller than $4\,R_{\oplus}$. The dot size is proportional to the planet radius, with super-Earths in brown and sub-Neptunes in blue. The thin vertical lines indicate the maximum orbital period for which the smallest planet in the system could have been detected with a SNR exceeding the threshold of 7. (No vertical line was drawn when this maximum period was more than 300 days.)
} 
\label{fig: Kepler multis system lineup}
\end{figure}

\newpage
\section{Peas-in-a-Pod and the Radius Valley}
\label{sec: Peas-in-a-Pod and the Radius Valley}

We categorize a given planet as either a super-Earth (SE) or a sub-Neptune (SN) based on whether its radius is smaller or larger than the boundary of the valley in $P$--$R_p$ space, as defined by \cite{2018MNRAS.479.4786V},
\begin{equation}
\log_{10}(R_p/R_{\oplus})=m\log_{10}(P/\mathrm{days}) + a,
\end{equation}
with $m=-0.09^{+0.02}_{-0.04}$ and $a=0.37^{+0.04}_{-0.02}$. Of the 2883 planets, 1265 are SEs and 1618 are SNs. Of the 520 multi-planet systems, 118 contain only SEs, 155 contain only SNs, and 247 (roughly half) contain both SEs and SNs. Figure \ref{fig: Kepler multis system lineup} illustrates the orbital spacings and sizes
for the 48 systems with at least four transiting planets smaller than 4\,$R_{\oplus}$. This subset is small enough to fit into a single plot, although our subsequent analysis includes all of the multi-planet systems. 

Visual inspection of Figure \ref{fig: Kepler multis system lineup} gives the impression that intra-system uniformity of planet sizes is much stronger for planets within the same size class (i.e., considering only points of the same color). A striking example is KOI-1060 (Kepler-758), which has a pair of SEs with radii $1.82\,R_{\oplus}$ and $1.87\,R_{\oplus}$, and a pair of SNs with radii $3.19\,R_{\oplus}$ and $3.16\,R_{\oplus}$. 

\begin{figure}[t!]
\epsscale{1.2}
\plotone{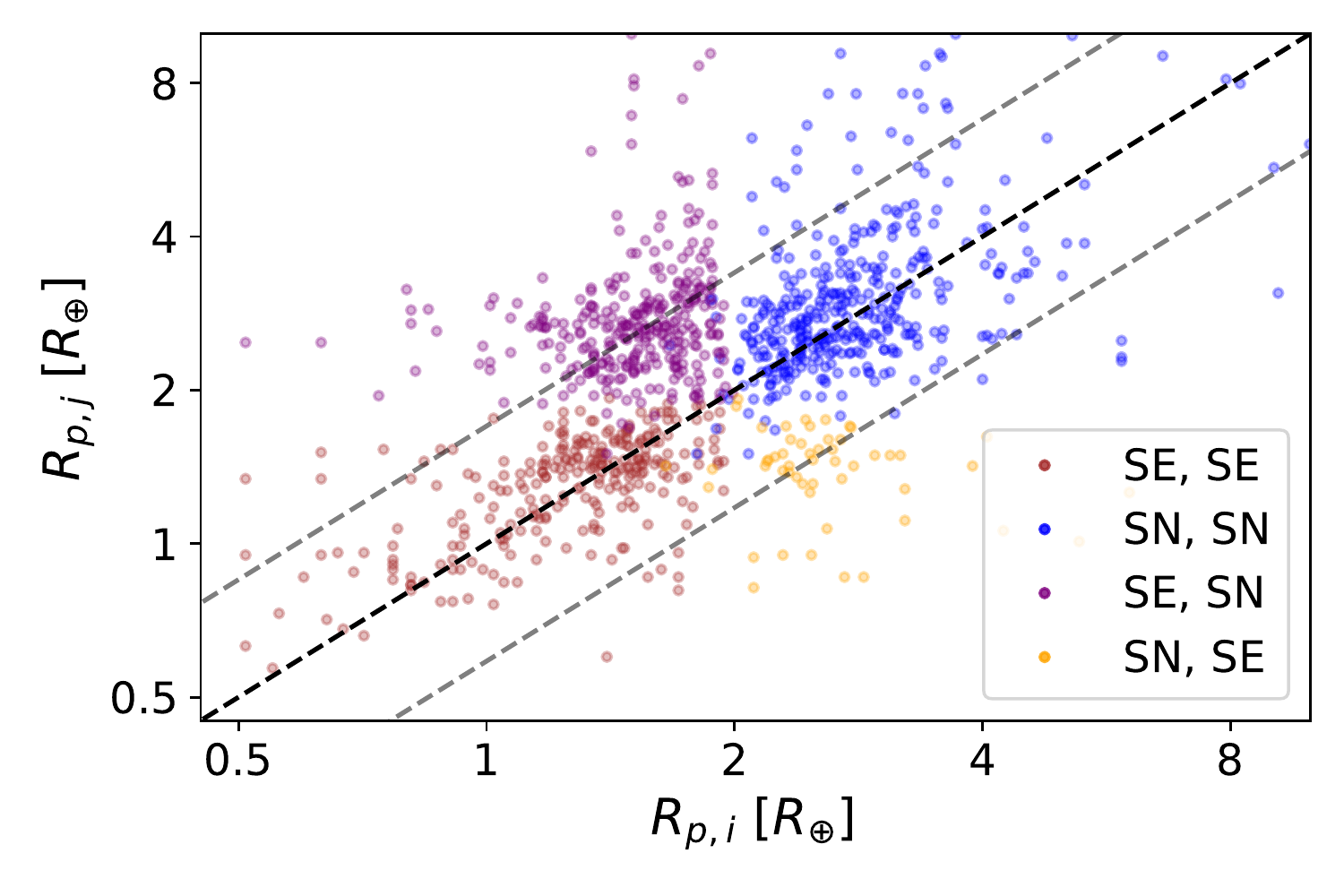}
\caption{Radii of pairs of planets within the same system, $R_{p,j}$ vs.\ $R_{p,i}$ where $P_i<P_j$, including non-adjacent pairs. The color of each
point conveys the size category of each planet in the pair. The dark dashed line is where $R_{p,j}=R_{p,i}$, and the lighter dashed lines are where $R_{p,j}$ is either $1.7\,R_{p,i}$ or $R_{p,i}/1.7$.} 
\label{fig: Rpj vs Rpi}
\end{figure}

\begin{figure*}
\centering
\epsscale{1.2}
\plotone{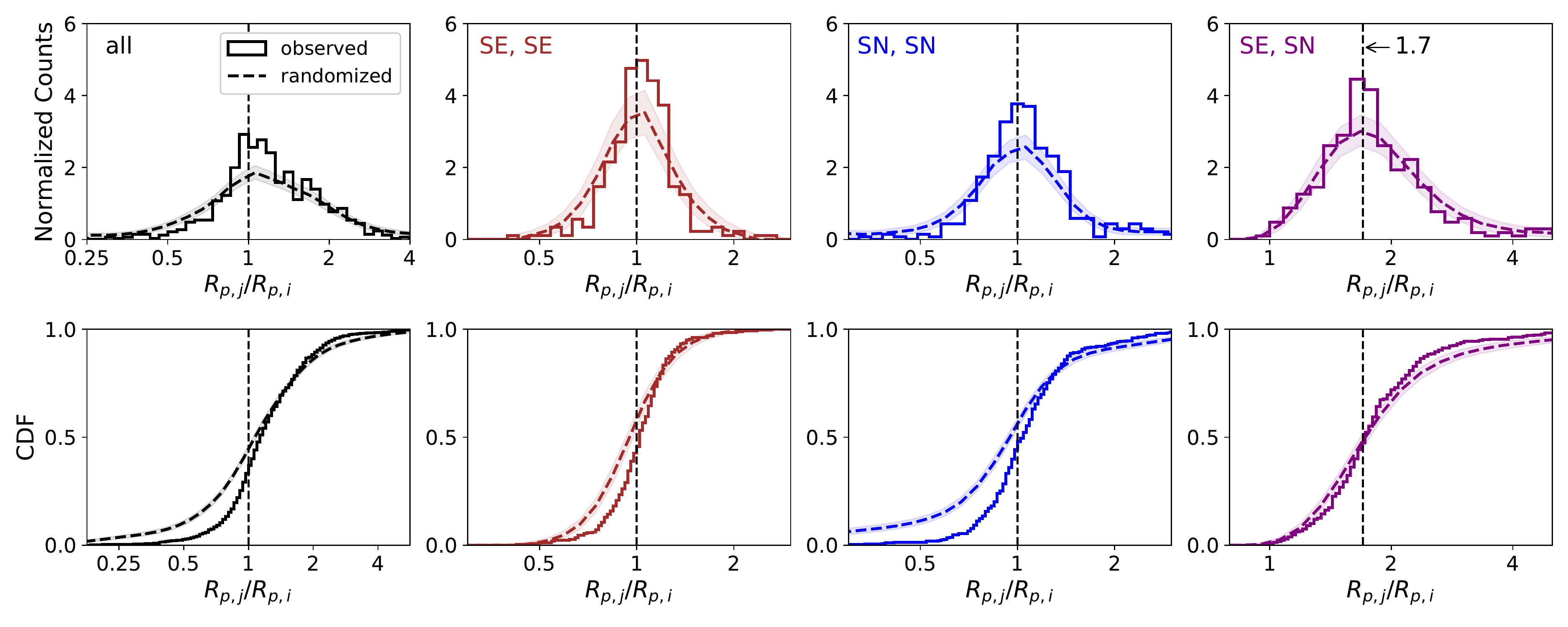}
\caption{Normalized histograms (top panels) and cumulative distributions (bottom panels) of the radius ratios of pairs of planets within the same system, $R_{p,j}/R_{p,i}$ where $P_i<P_j$. The distributions are split by pair type, with the columns corresponding to (1) all pairs; (2) SE,~SE; (3) SN,~SN; and (4) SE,~SN from left to right. The solid line corresponds to the observed data, while the dashed line and banding correspond to the median and 16th and 84th percentiles of the randomized trials. } 
\label{fig: radius ratio distributions}
\end{figure*}

For a quantitative analysis, we consider ratios between the radii of pairs of planets in the same system, $R_{p,j}$ vs. $R_{p,i}$ where $P_i<P_j$. To reduce asymmetries due to detection bias, we only include a planet pair in the set of radius ratios if it passes a ``swap test'', wherein both planets would still be detectable if they swapped periods \citep{2018AJ....155...48W}. In practice, this requires that the inner planet must still be detectable when assigned the longer period, $\mathrm{SNR}_{i}(P_i/P_j)^{1/3}>7$. Figure \ref{fig: Rpj vs Rpi} shows the radii of pairs of planets within the same system, including non-adjacent pairs. The points are colored by the four possible period-ordered types of pairs:
(1)~SE,~SE; (2)~SN,~SN; (3)~SE,~SN; and (4)~SN,~SE.
The count and fraction of the four pair types are 265 (26\%), 379 (38\%), 310 (31\%), and 52 (5\%), respectively. Thus, for 64\% of planet pairs, both planets are within the same size class, which helps to explain why intra-system uniformity was detectable in previous studies that did not split the planets into two categories.

As expected, Figure \ref{fig: Rpj vs Rpi} shows that the SE,~SE and SN,~SN pairs are much more clustered along the identity line -- indicating more uniform radii -- than the distribution as a whole. This is an inevitable consequence of restricting the size range of the planets before performing the comparisons.  More interesting is the quantification of the strength of the uniformity relative to the expectation of the null hypothesis, in which the radius ratios have the same distribution as would be obtained by drawing planets at random from the population \citep{2017ApJ...849L..33M, 2018AJ....155...48W, 2020ApJ...893L...1W}.

Null hypothesis testing is crucial because it indicates whether the uniformity trends are physical or the result of detection biases. Recently, there have been conflicting interpretations about the role of detection biases in sculpting the peas-in-a-pod patterns as a whole \citep{2020AJ....159..188Z, 2020AJ....160..160M}. However, comprehensive forward models \citep[e.g.][]{2019MNRAS.490.4575H, 2020AJ....160..276H} and population synthesis models \citep[e.g.][]{2021arXiv210512745M} that account for detection biases find strong evidence for intra-system uniformity within the Kepler planet population. (See also \citealt{2020ApJ...893L...1W}.) In the next section, we apply a simple model of transit detectability to evaluate planetary category-specific intra-system uniformity relative to random expectation.

\subsection{Significance assessment using randomized trials}

\setlength{\extrarowheight}{2pt}
\setlength\tabcolsep{3pt}
\begin{table*}[t!]
\centering
\caption{\textbf{Significance of intra-system radius uniformity and size ordering.}}
\begin{tabular}{c | c c c | c c c | c c c }
\hline
\hline
 & \multicolumn{3}{c|}{Pearson $r$ of $R_{p,j}$ vs. $R_{p,i}$} & \multicolumn{3}{c|}{MAD of $R_{p,j}/R_{p,i}$ distribution} & \multicolumn{3}{c}{fraction above $R_{p,j}/R_{p,i}=1$} \\
 & observed & randomized & significance & observed & randomized & significance & observed & randomized & significance \\
\hline
all pairs & $0.444$ & $0.030\pm0.033$ & $12.7\sigma$ & $0.417$ & $0.593\pm0.028$ & $-6.2\sigma$ & $0.660$ & $0.596\pm0.016$ & $4.0\sigma$ \\
SE,~SE & $0.570$ & $0.074\pm0.086$ & $5.8\sigma$ & $0.194$ & $0.258\pm0.027$ & $-2.4\sigma$ & $0.558$ & $0.499\pm0.041$ & $1.5\sigma$ \\
SN,~SN & $0.383$ & $0.002\pm0.048$ & $7.9\sigma$ & $0.264$ & $0.375\pm0.027$ & $-4.1\sigma$ & $0.549$ & $0.497\pm0.025$ & $2.1\sigma$ \\
SE,~SN & $0.103$ & $-0.011\pm0.063$ & $1.8\sigma$ & $0.432$ & $0.569\pm0.052$ & $-2.6\sigma$ & -- & -- & --  \\
\hline
\end{tabular}
\label{tab: significance calculations}
\end{table*}

In previous studies of intra-system uniformity, \cite{2018AJ....155...48W} and \cite{2017ApJ...849L..33M} constructed the random samples necessary for the null hypothesis testing by preserving the periods and multiplicities of each system but drawing the radii at random (with replacement) from the sample's entire distribution of planet radii. A drawback of this approach is that it does not preserve two prominent structures that are observed in $P$--$R_p$ space: the radius valley and the sub-Neptune desert (the paucity of sub-Neptunes in short-period orbits). Thus, some of the random systems are unlike any real system. Moreover, erasing the structure in $P$--$R_p$ space would lead to an overestimate of the strength of intra-system uniformity because the randomized systems would include SNs interior to SEs, a pattern which is rare in reality.
%We will quantify the impacts of this later.

We devised a procedure intended to preserve the important structures in $P$--$R_p$ space: 
\begin{enumerate}
\item Within each system, for each planet with radius $R_{p,i}$ and period $P_i$, draw a random radius $R_{p,i,\mathrm{rand}}$ from the subset of the full distribution with ${|\log_{10}P-\log_{10}P_i|<0.25}$ (with $P$ in days). 
\item Calculate the SNR for the newly drawn planet, $\mathrm{SNR}_{i,\mathrm{rand}}=\mathrm{SNR}_i(R_{p,i,\mathrm{rand}}/R_{p,i})^2$. If $\mathrm{SNR}_{i,\mathrm{rand}}<7$, redraw $R_{p,i,\mathrm{rand}}$ and repeat.
\item When computing radius ratios $R_{p,j,\mathrm{rand}}/R_{p,i,\mathrm{rand}}$ for pairs of planets in the same system, check whether the pair passes the swap test:  $\mathrm{SNR}_{i,\mathrm{rand}}(P_i/P_j)^{1/3}>7$. Otherwise, discard the radius ratio.  
\end{enumerate}
Using this procedure we created 1000 synthetic populations and corresponding distributions of radius ratios. Figure \ref{fig: P-Rp space} shows representative examples of $P$--$R_p$ diagrams of synthetic populations using the old and new methods.

Figure \ref{fig: radius ratio distributions} shows the radius ratio distributions of the observed and synthetic populations. We highlight three findings:

\textit{(1) Size uniformity.} All four observed distributions are more sharply peaked around the median than the distributions based on the synthetic systems. Table~\ref{tab: significance calculations} gives two measures of the statistical significance of intra-system uniformity: the Pearson $r$ correlation coefficient between $R_{p,i}$ and $R_{p,j}$ (see Figure \ref{fig: Rpj vs Rpi}) and the median absolute deviation from the median (MAD) of the $R_{p,j}/R_{p,i}$ distribution. The MAD quantifies the \textit{degree} of radius uniformity, which is about twice as strong for the SE,~SE and SN,~SN pairs than for all pairs (the MAD for the SE,~SE and SN,~SN pairs is $\sim0.2$ compared to $\sim0.4$ for all pairs). However, the \textit{significance} of the uniformity relative to the randomized trials is weaker for the SE,~SE and SN,~SN pairs because the comparisons are made within restricted radius ranges. 

\textit{(2) Size ordering.} 
One might expect longer-period planets to be larger on average because planets in wider orbits are
more likely to retain their atmospheres. Previous work has shown evidence for this effect in the overall planet population \citep{2013ApJ...763...41C, 2017ApJ...849L..33M, 2018MNRAS.473..784K, 2018AJ....155...48W}.
We can now check whether size ordering is also observed for the rocky cores and gas-enveloped planets. The CDFs of the SE,~SE and SN,~SN pairs in the bottom row of Figure~\ref{fig: radius ratio distributions} show only tentative evidence of size ordering; the observed distributions appear to be shifted to the right of the randomized distributions. Table \ref{tab: significance calculations} gives the fraction of the radius ratio distribution for which $R_{p,j}/R_{p,i}>1$. For the SE,~SE pairs, $56\%$ of the distribution has the larger planets on the outside, which is $1.5\sigma$ more than in the randomized trials. For the SN,~SN pairs, the corresponding numbers are $55\%$ and $2.1\sigma$.

\textit{(3) Peak at $\mathit{R_{p,\mathrm{SN}}/R_{p,\mathrm{SE}}\approx1.7}$.} The median of the SE,~SN radius ratio distribution occurs at $1.7^{+0.6}_{-0.3}$ (where the lower and upper ranges indicate 16th and 84th percentiles). This seems to be telling us that  ${R_p/R_{\mathrm{core}}\approx1.7}$, based on a simple physical argument. First,
we expect $M_p\approx\,M_{\mathrm{core}}$ because structural models show that the H/He dominated atmospheres of SNs constitute only a few percent of the planet's total mass \citep{2015ApJ...801...41R, 2015ApJ...806..183W}. Further
assuming that intra-system uniformity generally applies to planet mass as well as radius \citep{2017ApJ...849L..33M}, the uniformity in $M_p$ implies uniformity in $M_{\mathrm{core}}$, which in turn requires uniformity in $R_{\mathrm{core}}$ if the core compositions are similar in a given system. In this interpretation, SEs and SNs in the same system have approximately the same core size, and the characteristic radius ratio of 1.7 reflects the typical amount of gas of SNs. We will return to this point in Section~\ref{sec: Discussion}.

\section{Stellar Metallicity Dependence}
\label{sec: Stellar Metallicity Dependence}

\begin{figure}
\epsscale{1.}
\plotone{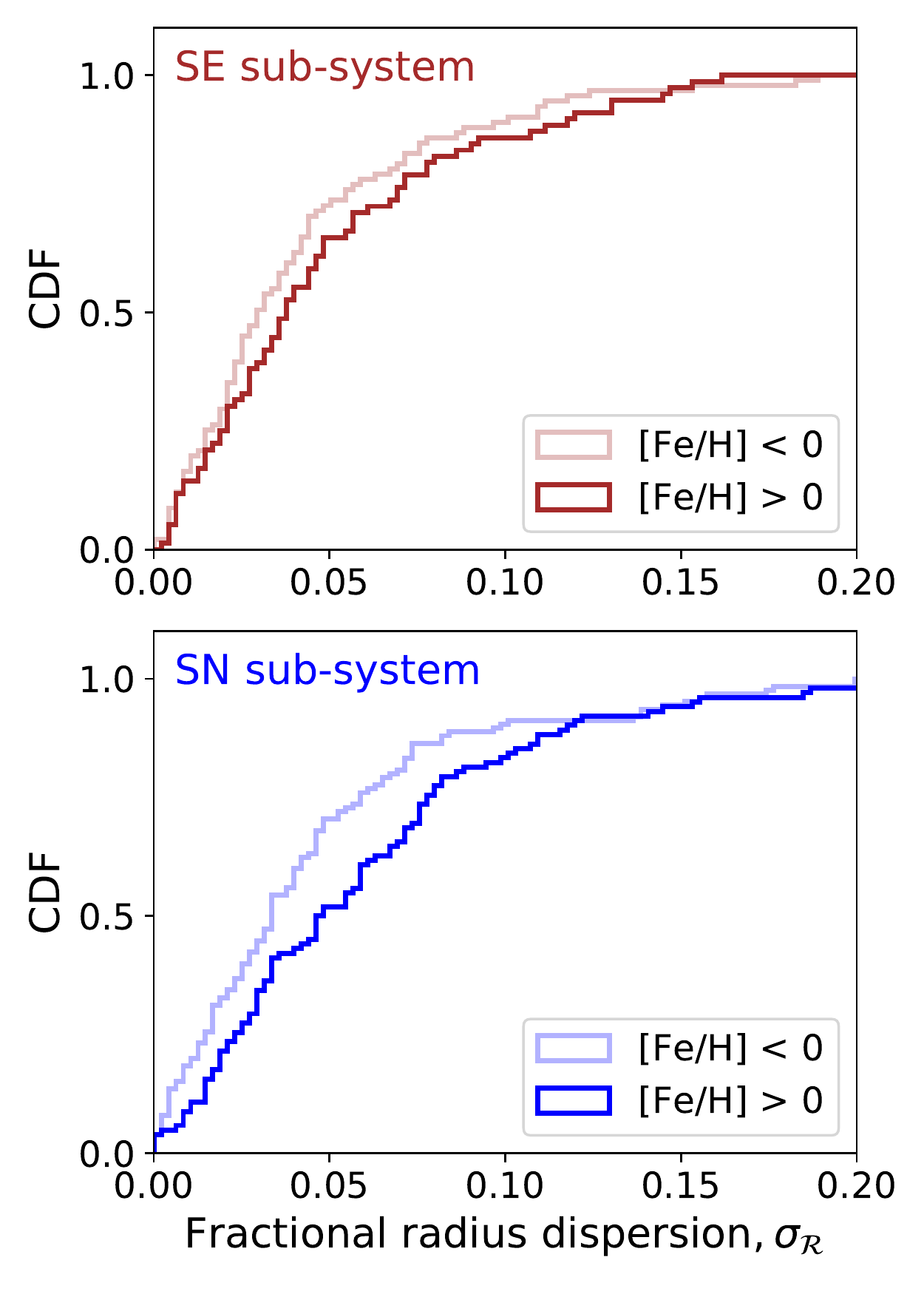}
\caption{Cumulative distributions of the fractional radius dispersion metric, split by size class and stellar metallicity. The top (bottom) panels corresponds to $\sigma_{\mathcal{R}}$ calculated for the subset of the planets in a system in the SE (SN)  size class. The lighter line colors correspond to systems with sub-solar metallicity, whereas the darker lines correspond to super-solar metallicity.} 
\label{fig: sigma_R CDFs}
\end{figure}

Previous work has shown that intra-system radius uniformity depends only weakly on stellar properties. \cite{2017ApJ...849L..33M} performed a multiple regression between planetary system characteristics and stellar properties and showed that the variation in metallicity [Fe/H] can explain only $\approx\,20\%$ of the variation of system-wide median planet sizes. Likewise, the variation in [Fe/H] could account for only $\approx\,10\%$ of intra-system size dispersion. To see whether these conclusions change when the planets are separated into distinct physical categories, we considered a metric for the intra-system fractional radius dispersion \citep{Weiss2021},
\begin{equation}
\sigma^2_{\mathcal R}={\rm Variance} \left\{ \mathrm{log_{10}}(R_{p,j}/R_\oplus) \right\},
\label{eq: sigma_R}
\end{equation}
where $j=1,2,...,N$ is indexed over planets in the system. We calculated $\sigma_{\mathcal\,R}$ for the SE sub-systems and the SN sub-systems and searched for any dependence on [Fe/H]. Figure \ref{fig: sigma_R CDFs} shows the results.

For the SE sub-systems, the Anderson-Darling test does not rule out the null hypothesis that the [Fe/H]~$<0$ and [Fe/H]~$>0$ sub-systems have the same distribution of $\sigma_{\mathcal R}$. For the SN sub-systems, though, the $p$-value is 0.005. Low-metallicity stars show slightly more uniformity in the sizes of their SNs than high-metallicity stars. This result agrees with the finding by \cite{2017ApJ...849L..33M} based on the concatenation of all planet types; we have shown here that the effect is more important for the SNs than for the SEs.

\section{Summary and Discussion}
\label{sec: Discussion}

Using a large sample of Kepler systems, we have explored the ``peas-in-a-pod pattern'' after splitting the ``peas'' into two physically distinct types of planets: super-Earths (SEs) and sub-Neptunes (SNs). The radius uniformity observed within SE,~SE and SN,~SN pairs is about twice as strong as one would infer without splitting the planets into these two categories. The MAD of the distribution of radius ratios ($R_{p,j}/R_{p,i}$ where $P_i<P_j$) for the SE,~SE and SN,~SN pairs is $\approx\,0.2$ (Table \ref{tab: significance calculations}), corresponding to a standard deviation of $\approx\,0.3$. Thus, to a reasonable approximation, $R_{p,j}/R_{p,i} \sim \mathcal{N}(\mu=1,\sigma=0.3)$. For SNs, the uniformity appears to be slightly enhanced for systems with lower stellar metallicity.

While radius uniformity is the dominant trend, there is $\sim$2-$\sigma$ evidence for size ordering even when considering only SE,~SE pairs or SN,~SN pairs. Size ordering of the planetary systems as a whole can be readily attributed to atmospheric loss for the highly irradiated planets \citep[e.g.][]{2013ApJ...776....2L} (i.e., SEs tend to be interior to SNs), but size ordering within each planet category may require a separate explanation.

In systems with both types of planets, the SN tends to be $1.7^{+0.6}_{-0.3}$ times larger than the SE. We argued that this implies $R_p\approx1.7\,R_{\mathrm{core}}$ for sub-Neptunes. The radius fitting formula from \cite{2019ApJ...886...72M} (see also \citealt{2014ApJ...792....1L, 2016ApJ...831..180C}) indicates that $R_p/R_{\mathrm{core}}\approx1.7$ corresponds to an envelope mass fraction ($f_{\mathrm{env}}\equiv\,M_{\mathrm{env}}/M_p$) of $f_{\mathrm{env}}\approx4.9\%$ for $M_p\approx10\,M_{\oplus}$ and $f_{\mathrm{env}}\approx2.6\%$ for $M_p\approx5\,M_{\oplus}$. The observation that $R_p\approx1.7\,R_{\mathrm{core}}$ is broadly consistent with both photoevaporation \citep[e.g.][]{2017ApJ...847...29O} and core-powered mass loss models \citep[e.g.][]{2019MNRAS.487...24G} and seems unlikely to help distinguish between them; however, this would benefit from further study.

Theoretical investigations of intra-system uniformity suggest that it can be explained by system-to-system variations in protoplanetary disk properties, combined with characteristic mass scales that emerge during planet conglomeration processes such as pebble accretion \citep{Weiss2021, 2021arXiv210512745M}. The mass scales depend on factors such as the disk scale height, solid surface density, and disk lifetime, thus yielding small variations within systems compared to those between systems. \cite{2019MNRAS.488.1446A} have also presented a simple model of planet formation, based on an energy optimization criterion,
in which uniform planet masses are the natural outcome. However, it is still unclear how specific aspects of the planet formation process map onto the quantitative \textit{degree} of uniformity that is observed. We hope our results will be useful for
future work that aims to understand the origins of both intra-system uniformity and diversity.

\section{Acknowledgements}
We thank the anonymous referee for their constructive comments, which helped improved this paper. S.C.M. was supported by NASA through the NASA Hubble Fellowship grant \#HST-HF2-51465 awarded by the Space Telescope Science Institute, which is operated by the Association of Universities for Research in Astronomy, Inc., for NASA, under contract NAS5-26555.

\newpage
\bibliographystyle{aasjournal}
\bibliography{main}

\end{document}